\documentclass[10pt]{iopart}
\usepackage{graphicx}
\usepackage{color}
\usepackage{epstopdf}
\usepackage{units}

\usepackage{braket}
\usepackage[usenames,dvipsnames,svgnames,table]{xcolor}
\usepackage{ulem}

\begin{document}
\title{Creation of entangled atomic states by an analogue of the Dynamical Casimir Effect}

\author{K.~Lange$^1$, J.~Peise$^1$, B.~L\"u{}cke$^1$, T.~Gruber$^2$, A.~Sala$^3$, A.~Polls$^{4,5}$, W.~Ertmer$^1$, B.~Juli\'a-D\'iaz$^{4,5}$, L.~Santos$^2$, C.~Klempt$^{1}$}

\address{$^1$Institut f\"ur Quantenoptik, Leibniz Universit\"at Hannover, Welfengarten~1, D-30167~Hannover, Germany}
\address{$^2$ Institut f\"ur Theoretische Physik, Leibniz Universit\"at Hannover, Appelstra\ss{}e~2, D-30167~Hannover, Germany} 
\address{$^3$ Department of Physics, NTNU, Norwegian University of Science and Technology, 7491 Trondheim, Norway} 
\address{$^4$ Departament de F\'isica Qu\`antica i Astrof\'isica, Facultat de F\'isica, Universitat de Barcelona, Barcelona 08028, Spain}
\address{$^5$ Institut de Ci\`encies del Cosmos, Universitat de Barcelona, ICCUB, Mart\'i i Franqu\`es 1, Barcelona 08028, Spain}

\ead{lange@iqo.uni-hannover.de}

\begin{abstract}
If the boundary conditions of the quantum vacuum are changed in time, quantum field theory predicts that real, observable particles can be created in the initially empty modes.
Here, we realize this effect by changing the boundary conditions of a spinor Bose-Einstein condensate, which yields a population of initially unoccupied spatial and spin excitations.
We prove that the excitations are created as entangled pairs by certifying continuous-variable entanglement within the many-particle output state.\\
\\
\end{abstract}

\maketitle
\ioptwocol

\section{Introduction}
\begin{figure}[htb!]
	\centering
	\includegraphics[width=.46\textwidth]{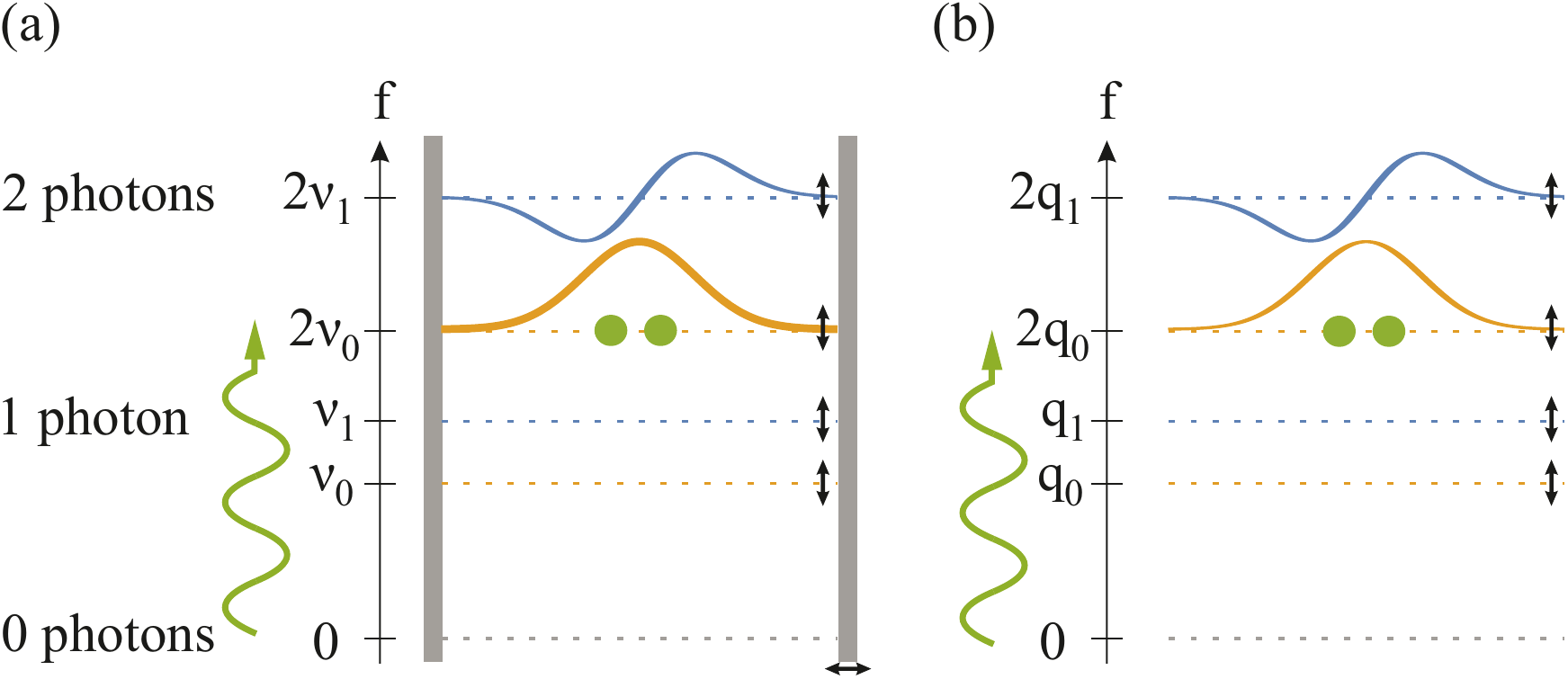}
		\caption{
		\label{fig1}		
		(a) Schematic principle of the originally proposed Dynamical Casimir effect. 
		The system includes two perfectly conducting plates which allow for a discrete spectrum of modes which may be populated with photons. 
		Changing the distance of plates non-adiabatically will result in an energetic shift of the light modes (black arrows). 
		If the frequency of the modulation (green arrow) is resonant to a certain mode, a pair of virtual photons is turned into real photons (green circles). 
		(b) Our analogue setup: 
		By manipulating the energy difference of the Zeeman levels, we can excite spins out of our spinor Bose-Einstein condensate into initially empty modes.
		}
\end{figure}

The quantized vacuum contains pairs of virtual quanta in all available modes of the physical system.
The static Casimir effect \cite{Casimir1948} is a measurable consequence of this process. 
In the original setting, two mirrors experience an attractive interaction due to the reduced number of electromagnetic modes in the volume between them.
First attempted in 1956 \cite{Lifshitz1956}, the Casimir force was precisely measured between a plane and a sphere in the late 1990s \cite{Lamoreaux1997, Mohideen1998} and between parallel plates in 2002 \cite{Bressi2002}.
If the mode density is not varied in space but in time, for example by a modulation of the mirror's distance \cite{Moore1970} or by changing the refractive index \cite{Yablonovitch1989}, the virtual quanta can be turned into macroscopic numbers of real excitations (see Fig.~\ref{fig1}a).
This so-called Dynamical Casimir effect (DCE) \cite{Schwinger1992} has been observed in the microwave regime \cite{Wilson2011,Laehteenmaeki2013}.
The excitations are created by nonadiabatic changes of the boundary conditions \cite{Dodonov2010,Dalvit2011}.
Moreover, this process creates pairs, which leads to the generation of entangled many-particle states out of the vacuum.

Bose-Einstein condensates (BECs) of diluted gases offer the possibility to study this effect either by modulating the magnetic field \cite{Saito2008} or the atomic scattering length \cite{Carusotto2010}.
In this case, the atomic BEC acts as a background field out of which atoms can be transferred to different, initially unoccupied modes.
These unpopulated modes represent the empty electromagnetic modes of the original setting.
These modes can be excited with either spatial (scalar Bogoliubov modes) or spin excitations. 
The external modulation must now alter the energy of these unpopulated modes, which can be realized by a modulation of the external trapping potential or the magnetic field.
Both schemes have been realized experimentally by driving spatial excitations~\cite{Jaskula2012} or spin excitations~\cite{Hoang2016}.
In Ref.~\cite{Jaskula2012}, the effect has been demonstrated for the first time, but the pure quantum character could not be proved.
While Ref.~\cite{Hoang2016} demonstrates the suppression of thermal excitations and a squeezing of spin-nematic observables, a proof that the DCE creates entanglement is missing.

In this article, we demonstrate the excitation of both spin and spatial degrees of freedom in a spinor BEC by the DCE.
We prove that the spin excitations are created in the form of entangled pairs by a violation of a continuous-variable entanglement criterion~\cite{Simon2000,Duan2000a}.
The experimental data is supported by a theoretical description of the system in terms of a numerical Bogoliubov analysis.

\section{Theoretical description}
\label{theory}
We consider an $F=1$ spinor BEC initially prepared in $m=0$. The system is described by the bosonic operators $\hat\psi_{m}(\vec r)$ that annihilate atoms with spin $m$ at position $\vec r$. 
Using Bogoliubov's approximation $\hat\psi_0(\vec r)=\psi_0(\vec r)+\delta{\hat\psi_0(\vec r)}$, where $\psi_0(\vec r)$ is the mean-field, and ${\delta\hat\psi_0(\vec r)}$ denote scalar fluctuations of the $m=0$ condensate. 
Up to second-order in the fluctuations, the spin fluctuations ${\hat\psi_{\pm 1}({\vec r})}$ decouple from the scalar fluctuations, and are given by the Hamiltonian:
\begin{eqnarray}
\hat H_1&=&\sum_{m=\pm1 }\int d^3 r\, \hat \psi_m^\dag (\vec r) \left ( \hat H_{\rm eff}+q \right )  {\hat\psi}_m(\vec r) \nonumber \\
&+& U_1\int d^3r\, n_0(\vec r)\left (\hat\psi_1^\dag(\vec r) \hat\psi_{-1}^\dag(\vec r) + \mathrm {H.c.} \right )
\label{eq:H1}
\end{eqnarray}
where $\hat H_{\rm eff}=-{\hbar^2\nabla^2}/{2M}+V(\vec r)+(U_0+U_1)n_0({\vec r})-\mu$, with $M$ the atomic mass, $n_0(\vec r)=|\psi_0(\vec r)|^2$, 
$V({\vec r})$ the external trapping potential, $\mu$ the chemical potential, $U_0=(g_0+2g_2)/3$, $U_1=(g_2-g_0)/3$, and $g_F=4\pi{\hbar^2 a_F}/{M}$, with $a_F$ the scattering-length for the 
collisional channel with total spin $F$. In Eq.~(\ref{eq:H1}), $q$ denotes the quadratic Zeeman energy~(QZE) term.
This energy may be externally modified using microwave fields. 
In particular, $q$ may be modified in time, which is a key feature in the following (see Fig.~\ref{fig1}b).

\subsection{Homogeneous case}

The connection between the spin dynamics in the presence of a time-dependent QZE and the DCE becomes particularly evident 
when considering a homogeneous BEC, i.e. $V({\vec r})=0$. In that case, $n_0({\vec r})=n$ is a constant and $\mu=U_0n$.
With this Eq.~(\ref{eq:H1}) may be re-written in momentum, $k$, space:
\begin{eqnarray}
\hat H_1&=&\sum_k  \bigg[  \sum_{m=\pm 1}\left ( \frac{\hbar^2k^2}{2M}+nU_1+q \right ) \hat a_{k,m}^\dag  \hat a_{k,m}  \nonumber \\
&+& U_1{n}\left ( \hat a_{k,1}^\dag {\hat a_{-k,-1}^\dag} + \mathrm{H.c.} \right ) \bigg],
\label{eq:H1}
\end{eqnarray}
where $\hat a_{k,m}$ denotes the bosonic operator for particles with spin $m$ and momentum $k$.
Employing the operators ${\hat a_{k,\pm}}=(\hat a_{{k,}1}\pm\hat a_{{k,}-1})/\sqrt{2}$, we introduce the Bogoliubov transformation $\hat a_{k,\pm}=\cosh \alpha_{k,\pm} \hat b_{k,\pm} + \sinh \alpha_{k,\pm}\hat b_{-k,\pm}^\dag$, with {$\sinh 2\alpha_{k,\pm}=\mp U_1 n/\xi(k)$}, where \\
$\xi(k)=\sqrt{\left (\hbar^2k^2/2M+q\right )\left (\hbar^2k^2/2M+q + 2 U_1n\right )}$ is the Bogoliubov spectrum of spin excitations. Using this transformation\\
$\hat H_1=\sum_k \xi(k) \left (\hat b_{k,+}^\dag \hat b_{k,+} + \hat b_{k,-}^\dag \hat b_{k,-} \right )$.

For the case of a sudden quench of the QZE from an initial value $q_i$ to a final one $q_f$, we introduce the Bogoliubov modes  
$\hat a_{k,\pm}=\cosh \alpha_{k,\pm} \hat b_{k,\pm} + \sinh \alpha_{k,\pm}\hat b_{-k,\pm}^\dag$, evaluated for $q_i$, and $\hat a_{k,\pm}=\cosh \tilde\alpha_{k,\pm} \hat c_{k,\pm} + \sinh \tilde\alpha_{k,\pm}\hat c_{-k,\pm}^\dag$, evaluated for $q_f$.  
The initial Bogoliubov modes fulfill the vacuum statistics $\langle  \hat b_{k,\pm}^\dag  \hat b_{k,\pm}\rangle =0$. 
When quenching q, the initial Bogoliubov modes project into the new ones: ${\hat c_{k,\pm}}=\cosh\Delta\alpha_{k,\pm} {\hat b_{k,\pm}} - \sinh\Delta\alpha_{k,\pm} {\hat b_{-k,\pm}^\dag}$, with $\Delta\alpha_{k,\pm}=\tilde\alpha_{k,\pm}-\alpha_{k,\pm}$.
As a result, as for the traditional Casimir effect, the quench of the QZE results right after the quench in non-zero occupations of the new Bogoliubov modes, $\langle \hat c_{k,\pm}^\dag (0) \hat c_{k,\pm}(0) \rangle= \sinh^2\Delta\alpha_{k,\pm}$. 
In addition, $\langle {\hat c_{k,\pm}(0) \hat c_{-k,\pm}(0)}\rangle  = -\frac{1}{2}\sinh 2\Delta\alpha_{k,\pm}$. 
After the quench, the evolution of the Bogoliubov modes is trivial $\hat c_{k,\pm}(t)=e^{-i\xi_f(k)t/\hbar} \hat c_{k,\pm}(0)$, where $\xi_f(k)$ is calculated for $q_f$.

This occupation of the spin Bogoliubov modes results in the creation of particles in $\ket{F,m}$=$\ket{1,\pm1}$. Indeed, using the 
relation between $\hat a_{k,\pm}$ and $\hat c_{k,\pm}$, we may obtain the population in $\ket{1,1}$, $n_{k,1}(t)=\langle \hat a_{k,1}^\dag \hat a_{k,1} \rangle$, which is 
at any time equal to that in $\ket{1,-1}$:
\begin{eqnarray}
n_{k,1}(t) &=&\cosh^2\tilde\alpha_{k,\pm}\sinh^2\Delta\alpha_{k,\pm}\nonumber\\
&+&\sinh^2\tilde\alpha_{k,\pm}\cosh^2\Delta\alpha_{k,\pm} \nonumber \\
&-&\frac{1}{2}\cos\left ( 2 \xi(k)t/\hbar \right ) \sinh2\tilde\alpha_{k,\pm}\sinh 2\Delta\alpha_{k,\pm}.
\label{eq:quench1}
\end{eqnarray}
Assuming a large $q_i\gg nU_1$, we may approximate $\alpha_{k,\pm}\simeq 0$, and hence
\begin{equation}
n_{k, 1}(t)=\left ( 1-\cos \left (\frac{2\xi_f(k)t}{\hbar}\right ) \right ) \left ( \frac{U_1n}{\xi_f(k)}\right )^2.
\label{eq:quench2}
\end{equation}
This creation of particles in the levels $\ket{1,\pm 1}$ constitutes the spin analogue of the recently reported Sakharov oscillations observed in scalar BECs when quenching the interactions \cite{Hung2013}. 

On the other hand, if $q(t)$ is periodically modulated in time, the resulting spinor dynamics resembles the DCE.
The Heisenberg equations for the Bogoliubov modes are of the form:
\begin{equation}
\frac{d}{dt}\hat b_{k,\pm}{(t)}=-\mathrm{i} f_k \hat b_{k,\pm}{(t)} \pm g_k{(t)} \hat b_{-k,\pm}^\dag{(t)},
\end{equation}
where $f_k\equiv {\xi(k)}/{\hbar}$ and $g_k{(t)}=- {\dot q(t)} {U_1n}/{2\xi(k)^2}$.
We introduce the expected values 
$P_{k{,\pm}}{(t)}\equiv \langle \hat b_{k,\pm}^\dag {(t)}\hat b_{k,\pm} {(t)}\rangle$,
$S_{k{,\pm}}{(t)}\equiv \pm \langle \hat b_{k,\pm}^\dag{(t)} \hat b_{-k,\pm}^\dag{(t)} \rangle + \mathrm{H.c.}$, and 
$A_{k{,\pm}}{(t)}\equiv \pm {\mathrm i} \left ( \langle \hat b_{k,\pm}^\dag{(t)} \hat b_{-k,\pm}^\dag {(t)}\rangle - \mathrm{H.c.} \right )$. 
{These two sets of equations can be summarized into one set by defining $P_k=P_{k,\pm}$, $S_k=\pm S_{k,\pm}$ and $A_k=\pm A_{k,\pm}$.}
The dynamics of these expected values is given by the equations:
\begin{eqnarray}
\dot  P_{k}{(t)}&=&g_k{(t)} S_{k}{(t)}, \\
\dot  S_{k}{(t)}&=&4g_k{(t)} P_{k}{(t)}+2g_k{(t)}+2f_k A_{k}{(t)}, \\
\dot  A_{k}{(t)}&=&-2f_k S_{k}{(t)},
\end{eqnarray}

The population in $m=\pm 1$ is given by:
\begin{eqnarray}
n_{k,\pm 1}(t)&=&\frac{\hbar^2k^2/2M+q{(t)}+nU_1}{\xi(k)}\left (P_{k}(t)+\frac{1}{2}\right )\nonumber\\
&-&\frac{1}{2}-\frac{U_1n}{\xi(k)}S_{k}(t),
\end{eqnarray}
which generalizes Eqs.~(\ref{eq:quench1}) and (\ref{eq:quench2}). Hence, as for the quench, the time dependent QZE results in a DCE, where the number of particles may be significantly enhanced employing a periodically-modulated $q(t)$ 
with a frequency matching resonantly one half of a Bogoliubov mode.

\subsection{Trapped case}
\label{trappedcase}

The analysis of the experimental realization of the Casimir effect demands a careful consideration of the trapping potential. 
In order to determine $\hat H_{\rm eff}$, we first obtain the initial density profile $n_0(\vec r)$ from the corresponding scalar 
Gross-Pitaevskii equation:
\begin{equation}
\mu \psi_0(\vec r)=\left [ -\frac{\hbar^2\nabla^2}{2M}+V(\vec r)+U_0 n_0(\vec r) \right ] \psi_0(\vec r).
\end{equation}
We then evaluate the eigenfunctions $\varphi_j({\vec r})$ of $\hat H_{\rm eff}$, such that $\hat H_{\rm eff} \varphi_j({\vec r}) = E_j \varphi_j({\vec r})$. 
Expressing $\hat \psi_m({\vec r})=\sum_j \varphi_j({\vec r}) \hat a_{j,m}$, we may re-express:
\begin{eqnarray}
\hat H_1&=&\sum_{j,m=\pm 1} (E_j+q) \hat a_{j,m}^\dag \hat a_{j,m} \nonumber\\
&+&U_1\sum_{i,j}\chi_{i,j}\left ( {\hat a_{i,1}^\dag \hat a_{j,-1}^\dag }+ \mathrm{H.c.} \right ), \label{eq:final_H}
\end{eqnarray}
with $\chi_{i,j}=\int d^3 r\, n_0(\vec r)\varphi_i(\vec r)\varphi_j(\vec r)$. 
For a sufficiently tight confinement, we may assume $\chi_{i,j\neq i}\ll \chi_{i,i}$, and $U_1\chi_{i,j\neq i}\ll |E_i-E_j|$~(this is indeed the case for our experimental parameters). 
In that case, $\hat H_1\simeq \sum_j \hat h_j$, with $\hat h_j = (E_j+{q})\sum_{m=\pm 1}\hat a_{j,m}^\dag \hat a_{j,m} + U_1 \chi_{jj} \left ( {\hat a_{j,1}^\dag\hat a_{j,-1}^\dag} + \mathrm{H.c.} \right )$. 
{We may then introduce the Bogoliubov transformation $\hat a_{j,\pm 1}=\cosh\alpha_{j,\pm} \hat b_{j,\pm}+\sinh\alpha_{j,\pm} \hat b_{j,\mp}^\dag$, with $\sinh 2\alpha_{j,\pm}=\mp U_1 \chi_{jj} / \xi_j$, where $\xi_j=\sqrt{(E_j+q)^2-(U_1\chi_{jj})^2}$ are the corresponding Bogoliubov energies.
 Then, apart from constants, $\hat h_j=\xi_j(\hat b_{j,+}^\dag \hat b_{j,+}+\hat b_{j,-}^\dag \hat b_{j,-})$}
{A maximal transfer rate to the excited spin states is thus reached for a large and imaginary Bogoliubov energy $\xi_j$, which is obtained for specific resonance conditions for $q$, where }
\begin{equation}
q_j=-E_j.
\label{eq:resonancecondition}
\end{equation}
We may proceed at this point as for the free-space case, obtaining the equations for the dynamics of the Bogoliubov modes: 
$\frac{d}{dt}\hat b_{j,\pm}{(t)}= \mathrm{i} f_j\hat b_{j,\pm}{(t)}+g_j{(t)}\hat b_{j,\mp}^\dag{(t)}$, with $f_j=\xi_j/\hbar$, and $g_j{(t)}=\dot q{(t)} U_1 \chi_{jj}/2\xi_j{(t)}$. 
We introduce $P_{j}{(t)}\equiv\langle\hat b_{j,\pm}^\dag{(t)}\hat b_{j,\pm} {(t)}\rangle$, $S_{j}{(t)}\equiv\langle\hat b_{j,+}^\dag{(t)}\hat b_{j,-}^\dag{(t)}\rangle  +\mathrm {c.c.}$, and $A_{j,+}{(t)}\equiv \mathrm{i} \left ( \langle\hat b_{j,+}^\dag{(t)}\hat b_{j,-}^\dag{(t)}\rangle  -\mathrm {c.c.}  \right )$.
Thus, the results for the trapped case resemble the free-space results by replacing the momentum states by eigenstates of the effective potential.

\section{Experimental observation} 

\begin{figure}[htb!]
	\centering
	\def\stackalignment{r}
\includegraphics[width=.46\textwidth]{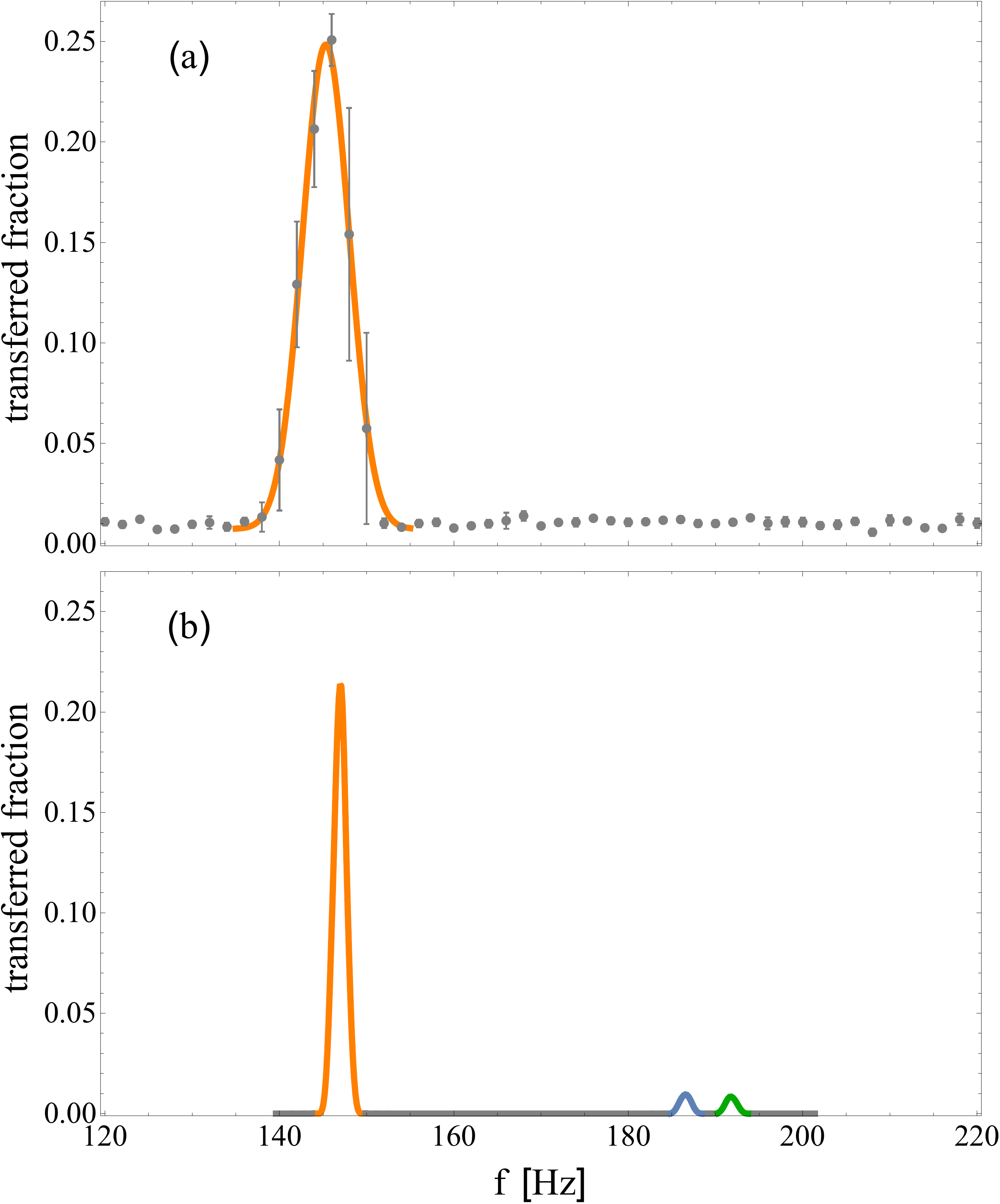}
	\caption{The relative number of atoms in the $m=\pm1$ levels is shown as function of the modulation frequency $f$ of the quadratic Zeeman energy. 
	(a) Experimental results with one clearly visible resonance corresponding to the ground mode. 
	The full orange line is a Gaussian fit, yielding a resonance frequency of $\unit[145]{Hz}$. 
	The maximum transfer corresponds to a excitation creation rate of  $\Omega/2\pi=\unit[1.06]{Hz}$. 
	(b) The theoretical calculations show one clearly visible resonance at $\unit[147]{Hz}$ with a corresponding excitation creation rate of $\Omega/2\pi=\unit[1.01]{Hz}$.
	 Both the position and the creation rates match the experimental results.
	Two smaller resonances are also visible at the frequencies $\unit[187]{Hz}$ and $\unit[192]{Hz}$, corresponding to spatially excited modes.}
	\label{fig2}	
\end{figure}

We employ almost pure $^{87}$Rb BECs in a crossed-beam optical dipole trap with trapping frequencies $\unit[2\pi\times(150,160,220)]{Hz}$.
The 22,000 atoms in the BEC are prepared in the hyperfine level $\ket{1,0}$.
At our applied magnetic field of $B = \unit[2.6] {G}$, the magnetic field-induced QZE is $q_{B}=\unit[487]{Hz}$.
Before initiating the dynamics, we empty the levels $\ket{1,\pm1}$ with two microwave pulses  from $\ket{1,\pm1}$ to $\ket{2,\pm2}$ followed by a light push resonant to the F=2 manifold to ensure that there are no excitations present in these levels.
In our experiments, we apply an effective shift of the QZE $q_d$ by a microwave dressing field that couples the levels $\ket{1, -1}$ and $\ket{2, -2}$.
Atoms are transferred from the level $\ket{1,0}$ to the levels $\ket{1,\pm 1}$ in the trap's ground mode if the sum of dressing field and the magnetic-field-induced QZE matches the resonance condition~(\ref{eq:resonancecondition}), $q\equiv q_B+q_d=q_0$~\cite{Klempt2009,Scherer2010,Scherer2013}.
Fig.~\ref{fig3}a shows this transition from the stable into the unstable region with the according resonance in the number of transferred atoms at the boundary of these regions (in orange).
There are further resonances (blue in Fig.~\ref{fig3}a) at $q=q_j<q_0$, when the difference $q_0-q_j$ is approximately equal to the energy difference $E_j-E_0$ to the $j$th excited mode of the effective potential.
Otherwise, the BEC remains in the state $\ket{1,0}$ and no atoms are transferred.
In our experiments, the first two excited spatial modes are seen as one resonance, because two trap frequencies are close to degeneracy.
Nevertheless, as shown with the absorption images, they can be individually addressed by choosing the correct QZE.

The analogue DCE is realized in the regime $q>q_0$, where the BEC is stable.
Here, the intensity of the microwave field is modulated sinusoidally, yielding a corresponding oscillation of the QZE.
If the frequency of the QZE oscillation is resonant to approximately twice the QZE difference to a specific resonance, $f= 2 (q-q_j)/h$, atoms are parametrically excited to the respective mode.
This process can be described as a parametric amplification of vacuum fluctuations in the $\ket{1,\pm 1}$ modes.
The number of the transferred atoms is detected by state-selective absorption imaging.
We will show that the amplification of vacuum fluctuations leads to measurable populations in the levels $\ket{1,\pm 1}$ in spin and spatial degrees of freedom and confirm the quantum origin of the dynamics by quantifying the created continuous-variable entanglement.

\subsection{Dynamical Casimir ground-mode resonance}
\label{DynamicalCasimirgroundresonance}
In our experiments, we observe the analogue DCE by setting the QZE to a value of $(q - q_0)/h = \unit[71]{Hz}$, far in the stable regime.
We modulate the QZE $q/h$ for $\unit[700]{ms}$ with an amplitude of $\unit[48]{Hz}$ by controlling the intensity of the microwave dressing field.
Figure~\ref{fig2}a shows the fraction of transferred atoms as a function of the modulation frequency $f$.
The data shows a resonance at $\unit[145]{Hz}$, which is approximately twice the QZE difference to the ground mode, $2 (q-q_0)/h=\unit[142]{Hz}$.
The data may be compared to the theoretical prediction in Fig.~\ref{fig2}b.
Here, the frequency of the ground-mode resonance at $\unit[147]{Hz}$ is in good agreement with the experimental results.
The difference between the theoretical prediction and twice the QZE difference $2 (q-q_0)/h$ may be explained by slight inaccuracies in the determination of the modulated QZE from dc measurements, as well as drifts and anharmonicities in the trapping potential.
On resonance, the transferred fraction of atoms follows an exponential growth.
We calculate a theoretical spin excitation rate of $\Omega/2\pi=\unit[1.01]{Hz}$ that matches well the experimental rate of $\Omega/2\pi=\unit[1.06]{Hz}$, as it is obtained from the maximally transferred fraction on resonance. 
In contrast to the theoretical calculations, the experimental resonance width of $\unit[2.7]{Hz}$ is four times larger than the width of the theoretical resonance of $\unit[0.7]{Hz}$.
This is a result of the varying total number of atoms in our BECs as discussed below.
The excited state resonances at the frequencies $\unit[187]{Hz}$ and $\unit[192]{Hz}$ are not visible in our experimental data.
We will address this issue in the next paragraph.

\subsection{Excited resonance}

\begin{figure}[htb!]
\centering
\includegraphics[width=.46\textwidth]{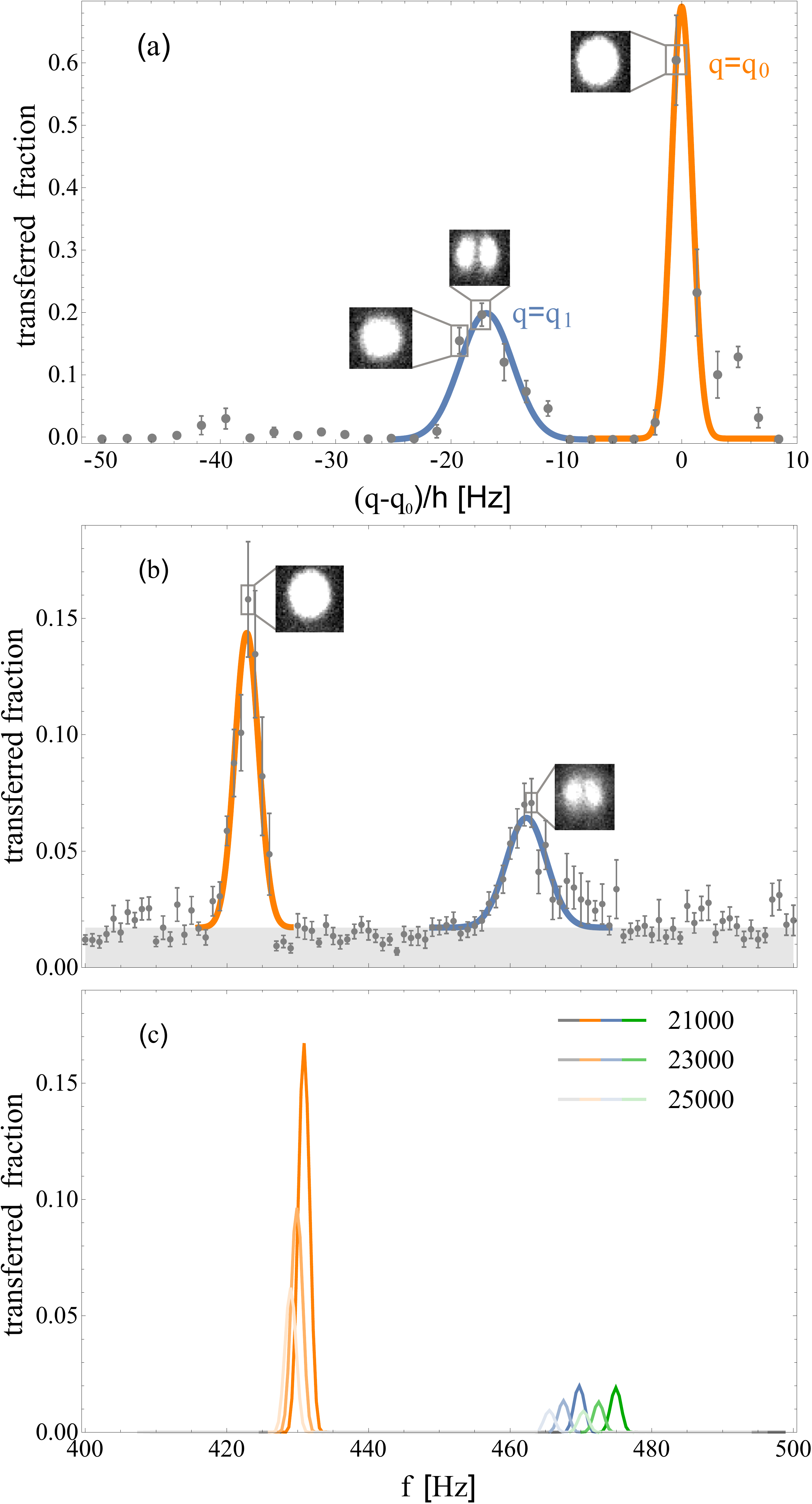}
\caption{(a) For specific values of the QZE $q=q_j$, we observe well resolved resonances in the unstable region of the BEC, where distinct spatial modes are populated via spin dynamics towards the levels $\ket{1,\pm 1}$. 
The spatial profile of the modes is observed in our absorption images (see inset).
The colored lines are Gaussian fits to guide the eye. 
(b) Parametric excitations populate distinct resonances by varying the modulation frequency of the quadratic Zeeman energy after generating seed atoms with collisional interactions.
The gray shaded area indicates the mean transfer due to spin-changing collisions.
Same colors indicate the same modes.
(c) Theoretically obtained resonances for different total numbers of atoms. 
The brightness corresponds to different total number of atoms and different colors indicate different modes.}
\label{fig3}	
\end{figure}

We further study the analogue DCE on an excited spatial mode.
As seen in Fig.~\ref{fig2}b, the creation rate of excitations is smaller and narrower for excited spatial modes due to the reduced mode overlap $\chi_{jj}$.
For such reduced spin excitation rates, our system is dominated by additional loss processes, such as atomic collisions that transfer the atoms from the excited trap mode to the ground mode.
As the exponential growth rate depends on the bosonic enhancement of initially transferred atoms, a significant loss can lead to a complete inhibition of the growth process.
Furthermore, fluctuations of the total number of atoms lead to a further suppression of the resonance, as discussed below.

To mitigate the influence of the loss processes, we enhance the creation rate on the first excited spatial mode by an initial transfer of seed atoms to the chosen mode.
Prior to our DCE protocol, we deliberately transfer seed atoms to the first excited spatial mode by choosing a resonant QZE $q=q_1$ (see Fig.~\ref{fig3}a).
The creation of seed atoms is facilitated by enabling spin-changing collisions via our microwave dressing for $\unit[150]{ms}$ on the first excited spatial mode at $(q-q_0)/h=\unit[-16.9]{Hz}$.
In the mean, $\unit[1.7]{\%}$ of the atoms are transferred to the excited spatial mode (gray shaded area in Fig.~\ref{fig3}b).

We further increase the signal by increasing the modulation amplitude.
We oscillate $q/h$ from $\unit[45]{Hz}$ to $\unit[363]{Hz}$ for a modulation time of $\unit[650]{ms}$. 
Due to a nonlinearity, the oscillation is slightly distorted from a pure sinusoidal shape and is centered around  $\unit[214]{Hz}$.
For these experimental parameters, we observe not only the population of the ground mode of the effective potential, but also of the seeded first excited mode (see Fig.~\ref{fig3}b).
The frequencies of the ground-mode and the excited-mode resonances are determined by Gaussian fits yielding $\unit[422]{Hz}$ and $\unit[462]{Hz}$.
The inset in Fig.~\ref{fig3}b shows the spatial mode profile of the excited atoms. 
Resonances corresponding to a certain mode display the respective spatial profile. 

The theoretical calculations in Fig.~\ref{fig3}c agree qualitatively with the experimental results. 
The results are displayed for three different total numbers of atoms.
The positions of the excited-mode resonances shift several resonance widths depending on the number of atoms.
This number-dependent shift of the narrow lines combined with our experimental fluctuations of the total atoms number of 1800 atoms presents a reason why we were unable to observe the excited resonances without seed atoms.
Furthermore, the theoretical results show a systematic shift to higher modulation frequencies compared to the experimental results. 
Again, this effect may be explained by slight inaccuracies in the determination of the modulated QZE from dc measurements, as well as drifts and anharmonicities in the trapping potential.
Our results show the parametric excitation of atoms into different spin and spatial modes by an analogue of the DCE.

\section{Entanglement characterization}

\begin{figure}[htb!]
\centering
\includegraphics[width=.46\textwidth]{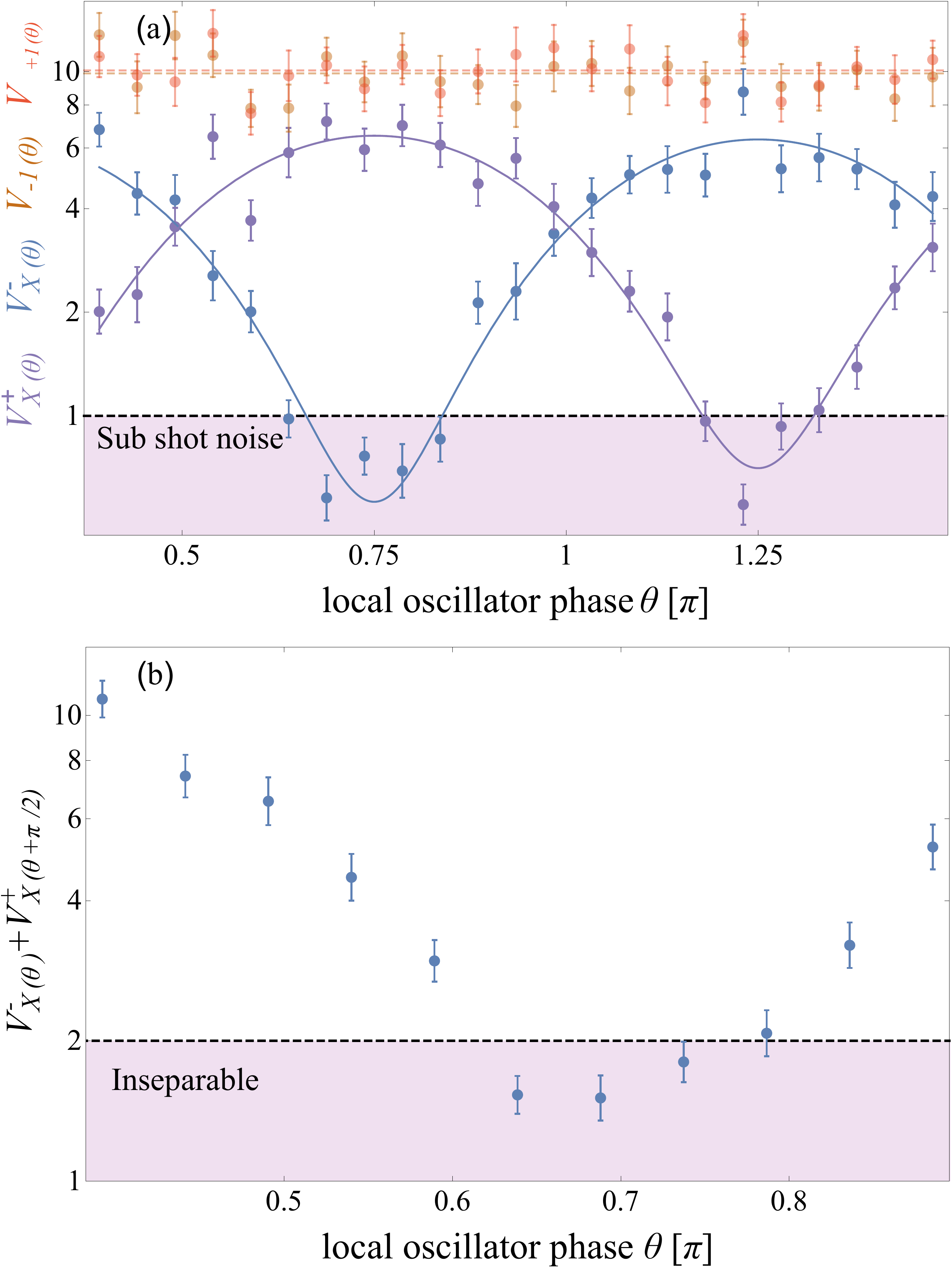}
\caption{(a) One-mode variances $V_{-1}$ and $V_{+1}$ and two-mode variances $V_d$ and $V_s$ as a function of the local oscillator phase $\theta$. The variances of the individual modes show no phase evolution and fluctuations greater than shot-noise. The two-mode variances are squeezed below shot-noise at $\theta=0.75\,\pi$ for $V_{d}$ and at $\theta=1.25\,\pi$ for $V_{s}$. (b) Inseparability parameter $V_d{(\theta)}+V_s{(\theta+\pi/2)}$ as a function of the local oscillator phase $\theta$. The dashed line indicates inseparability of the underlying quantum state. We violate this boundary by 2.3 standard deviations which proofs continuous-variable entanglement.}
\label{fig4}	
\end{figure}

In this section, we prove the quantum nature of DCE by demonstrating the quantum correlations between the excitations created in the two modes.
For these experiments, we employ the sequence of section \ref{DynamicalCasimirgroundresonance}, with a shorter modulation time of $\unit[110]{ms}$.
Following our previous work \cite{Peise2015a}, we demonstrate the quantum correlations between the quadratures of the levels $\ket{1,\pm1}$, defined as $\hat x_{\pm1}={1}/{\sqrt{2}}\,(\hat a^\dag_{\pm1}+\hat a_{\pm1})$ and $\hat p_{\pm1}={i}/{\sqrt{2}}\,(\hat a^\dag_{\pm1}-\hat a_{\pm1})$. 
In our experiments, we detect either the $x$ or the $p$ quadratures of both levels $\ket{1,\pm 1}$ by unbalanced atomic homodyne detection. 
Our BEC acts as the local oscillator for the homodyne detection.
A radio-frequency pulse couples $15\%$ of the local oscillator with the levels $\ket{1,\pm1}$. 
The local oscillator phase $\theta$ can be adjusted via a variable holding time with deactivated microwave dressing. 
For each holding time, we obtain a linear combination of both quadratures $\hat X_{\pm1}(\theta)=\hat x_{\pm1}\cos(\theta-\pi/4)+\hat p_{\pm1}\sin(\theta-\pi/4)$, with the corresponding variances $V_{\pm 1} \equiv \mathrm{Var}[\hat X_{\pm1}(\theta)]$.
For $\theta=3\pi/4$, the variance of the difference $V_d \equiv \mathrm{Var}[\hat X_{+1}(\theta)-\hat X_{-1}(\theta)]$ is squeezed, while for $\theta=5\pi/4$, the variance of the sum $V_s \equiv \mathrm{Var}[\hat X_{+1}(\theta)+\hat X_{-1}(\theta)]$ is squeezed.
Consequently, the local oscillator phases $3\pi/4$ and $5\pi/4$ can be associated with the $x$ and $p$ quadratures.
These two quadratures show sub-shot-noise fluctuations (blue and purple dots in Fig.~\ref{fig4}a), which indicates two-mode squeezing. 
Additionally, no phase dependence is visible for the quadrature correlations of the individual modes (red and orange dots in Fig.~\ref{fig4}a). 
As a consequence, there is no one-mode squeezing, which follows our predictions. 
We prove entanglement with the inseparability criterion $V_d + V_s < 2$ \cite{Simon2000,Duan2000a} for two collective atomic modes (see Fig.~\ref{fig4}b). 
The strongest violation is $V_d + V_s = 1.51\pm0.17$ proving entanglement with 2.9 standard deviations. 

\section{Conclusion}
In conclusion, we have demonstrated that spin dynamics in spinor BECs resembles the DCE.
We have observed the generation of atom pairs in initially empty excited states of the system by a resonant modulation of the energy of the excited states.
The created pairs carry entanglement, which we have proven by detecting the non-classical correlations between the quadratures.
This central finding unveils the deep connection between the Casimir Effect and the generation of non-classical states.
In the future, the parametric generation of entangled atom pairs can be employed as a versatile tool for the generation of entangled atomic ensembles.
The modulation method offers a fast initialization of the pair generation process compared to conventional methods, where the resonance conditions is reached by ramping the QZE to the unstable regime.

\ack{
{We acknowledge support from the Centre for Quantum Engineering and Space-Time Research (QUEST), Generalitat de Catalunya through grant 2014SGR-401, the Ministerio de Economia (Spain) through grant FIS2014- 54672-P and the Deutsche Forschungsgemeinschaft (DFG) through project KL2421/2-1, RTG 1729, CRC 1227 (DQ-mat), project A02.
}

\section*{References}

\bibliography{main}

\end{document}